# Studies of two-dimensional MoS$_2$ on enhancing the electrical performance of ultrathin copper films

*Tingting Shen[1,3\*], Daniel Valencia[2,4], Qingxiao Wang[5], Kuang-Chung Wang[2], Michael Povolotskyi[3],*

*Moon J. Kim[5], Gerhard Klimeck[2], Zhihong Chen[2,3], and Joerg Appenzeller[2,3]*

[1] Department of Physics and Astronomy, Purdue University, West Lafayette, IN, USA 47907

[2] Department of Electrical and Computer Engineering, Purdue University, West Lafayette, IN, USA 47907

[3] Birck Nanotechnology Center, Purdue University, West Lafayette, IN, USA 47907

[4] The Charles Stark Draper Laboratory Inc., Cambridge, MA, USA 02139

[5] Department of Materials Science and Engineering, University of Texas at Dallas, Richardson, TX, USA 75080

**Abstract:**

Copper nanowires are widely used as on-chip interconnects due to superior conductivity. However, with aggressive Cu interconnect scaling, the diffusive surface scattering of electrons drastically increases the electrical resistivity. In this work, we studied the electrical performance of Cu thin films on different materials. By comparing the thickness dependence of Cu films resistivity on MoS$_2$ and SiO$_2$, we demonstrated that two-dimensional MoS$_2$ can be used to enhance the electrical performance of ultrathin Cu films due to a partial specular surface scattering. By fitting the experimental data with the theoretical Fuchs–Sondheimer (FS) model, we claimed that the specularity parameter at the Cu/MoS$_2$ interface is p ≈ 0.4 in the temperature range 1.8K < T < 300K. Furthermore, first principle calculations based on the density functional theory (DFT) indicates that there are more localized states at the Cu/amorphous SiO$_2$ interface than the Cu/MoS$_2$ interface which is responsible for the higher resistivity in the Cu/SiO$_2$ heterostructure due to more severe electron scattering. Our results suggest that Cu/MoS$_2$ hybrid is a promising candidate structure for the future generations of CMOS interconnects.

**Introduction**

Copper is widely used as the interconnect material due to superior conductivity [1-4]. Along with the scaling of VLSI circuits, the scaling of the interconnects is also highly demanded. The scaling trends of the height and width of the interconnects are shown in the International Technology Roadmap for Semiconductors (ITRS) [4]. However, when the thickness of Cu film decreases down to the electron mean free path which is 40nm at room temperature [5], the electrical resistivity will increase significantly due to the increased electron scatterings from film surfaces [6-8] and grain boundaries [9-10]. This size effect impacts the time delay of the interconnects severely and represents a major challenge for the development of nanoelectronics [3-4]. To resolve this problem, novel Cu/barrier interfaces with specular rather than diffusive electron scattering need to be developed to achieve high-conductivity interconnects [11-12]. Although pristine atomically smooth Cu surface shows partial specular scattering [11, 13], there are many factors such as surface roughness [8], oxidation in ambient environment [10-12] or coatings of secondary materials [14] that will lead to completely diffusive surface scattering due to the randomization of the electron momentum in the current flow direction. Furthermore, interconnect system requires not only good electrical performance but also the capability to mitigate Cu diffusion into damascene structures. Conventional barrier materials such as Ta/TaN and TiN have been used to isolate Cu from the surrounding dielectrics. Since the resistivity of the conventional barrier materials are more than one order higher than that of Cu [15], the thickness of the barrier material needs to be reduced as much as possible to maximize the Cu volume for lower line resistance. But when the thickness of the conventional barrier material is scaled down below 3nm, the barrier cannot block Cu diffusion anymore. To achieve high performance integrated circuits, sub-1nm novel diffusion barriers which yield specular electron surface scattering are highly demanded.

2D layered materials have attracted intense research interest for the application in copper interconnect technology [12, 16-20] due to their ultra-thin body thickness. Recent studies show that atomic graphene not only has excellent performance in blocking Cu diffusion [19-20] but also can enhance the electrical and thermal conductivity of Cu [12]. R. Mehta *et al.* [12] reported a partial specular scattering of p = 0.23 at graphene coated Cu surfaces. On the other hand, two-dimensional layered semiconducting transition metal dichalcogenides (TMDs) like $MoS_2$ have also been indicated to be good Cu diffusion barrier materials [16]. However, there are few studies on the electrical properties of Cu/$MoS_2$ barrier hybrid [18]. In this letter, we studied the electrical performance of Cu thin films on different materials. Our experimental results show that for Cu films with the same thickness, Cu on $MoS_2$ always show much lower resistivity than Cu on $SiO_2$. Analyzing the relationship between Cu resistivity and thickness at different temperatures, we demonstrate that surface scattering is the main contribution to the total resistivity when Cu is thinner than 100nm, and the Cu/$MoS_2$ interface shows partial specular scattering with a temperature independent specularity p = 0.4 which is better than that reported in Cu/Ni [11] and Cu/graphene [12] structures. Furthermore, we studied the electronic properties of four different Cu surfaces: pure Cu, Cu/amorphous $SiO_2$, Cu/crystalized $SiO_2$ and Cu/$MoS_2$ by first principle calculations based on the density functional theory (DFT). It is found that: (1) the DOS of Cu/$MoS_2$ interface is similar with pure Cu surface; (2) the DOS of Cu/amorphous $SiO_2$ and Cu/crystalized $SiO_2$ is much higher than that of Cu surface. It is worth to mention that the states at the interfaces are localized which will trap electrons traversing near the interface. Upon subsequent release, the electron momentum will be randomized in the current direction. Thus, the Cu/$SiO_2$ heterostructure always show higher resistivity than the Cu/$MoS_2$ heterostructure since the high localized states at the Cu/$SiO_2$ interface caused complete inelastic surface scattering. Our results indicate Cu/$MoS_2$ hybrid has significantly improved the electrical performance of thin Cu films which is highly desirable for future generations of CMOS interconnects.

**Results and Discussion:**

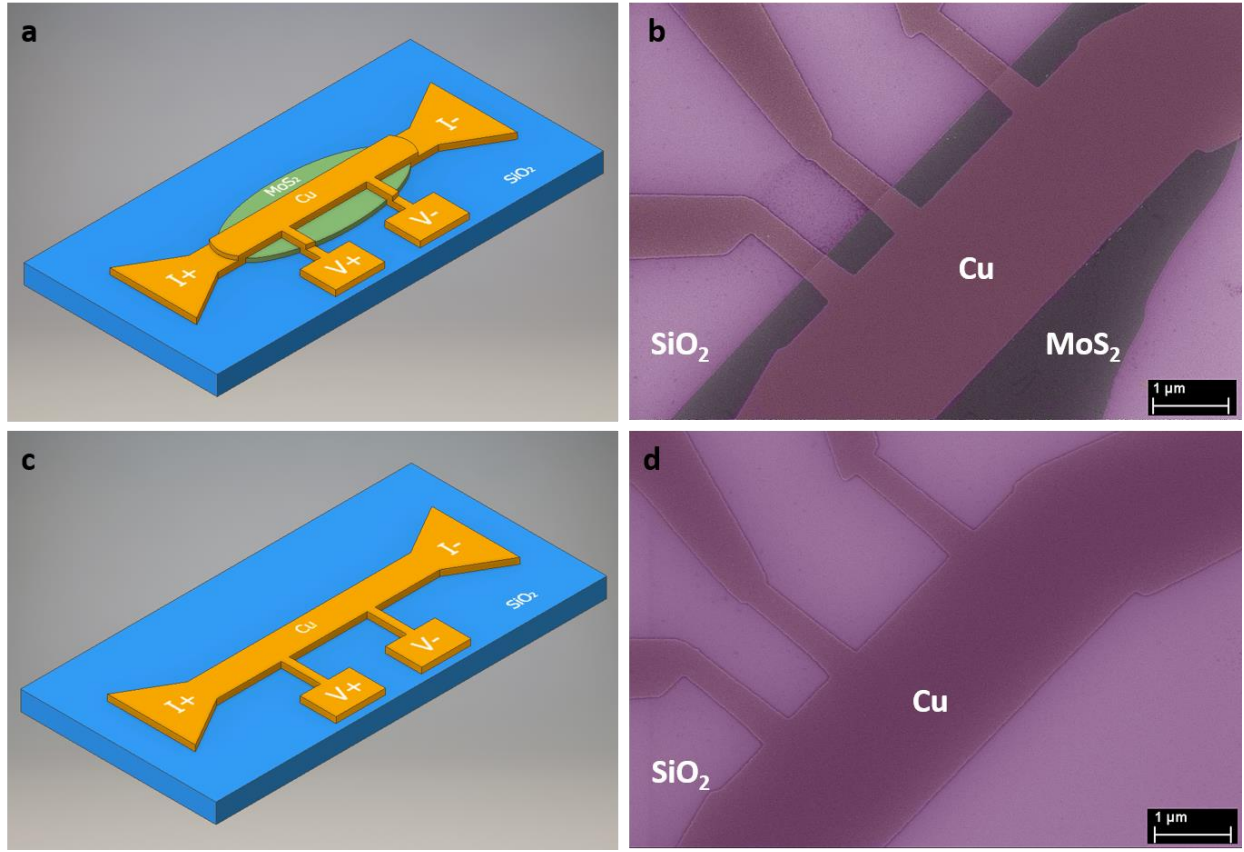

Figure 1. (a) Schematic diagram and (b) representative SEM image of a Cu on $MoS_2$ device. (c) Schematic diagram and (d) representative SEM image of a Cu on $SiO_2$ device.

We fabricated two types of devices, Cu on $MoS_2$ and Cu on $SiO_2$, to study the electrical performance of Cu thin films on different materials. Figure 1(a)-(d) show the schematic diagrams and representative SEM images of Cu/$MoS_2$ and Cu/$SiO_2$ devices respectively. To achieve fair Cu electrical performance comparison, these two types devices were patterned into structures with the same dimensions on the same Si/$SiO_2$ substrates as the representative SEM images show in Figure 1(b) and Figure 1(d). Cu thin films of different thickness were deposited using an e-beam evaporation system and the electrical resistance was measured by four-probe methods in a probe station set-up. The measurement geometry is shown in Figure 1(a) and Figure 1(c). Details of the fabrication are described in the "Methods" section.

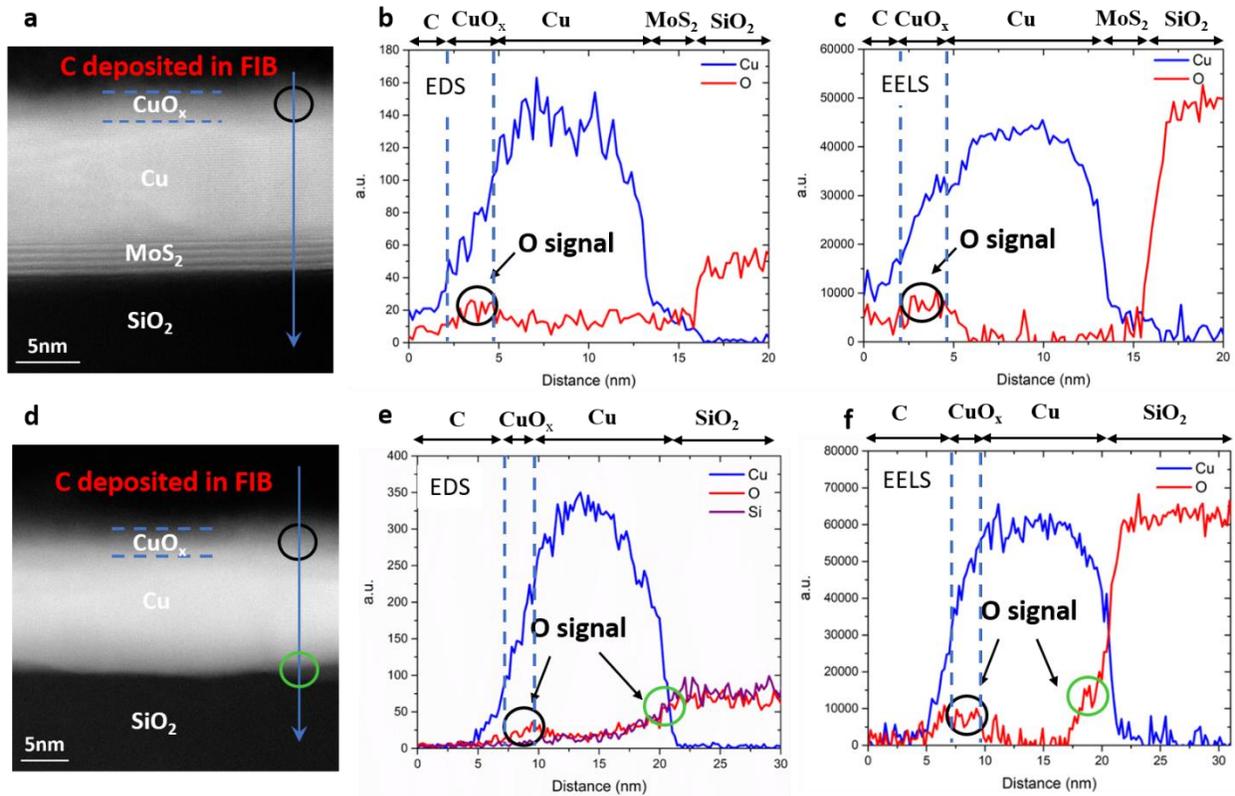

Figure 2. Cross-sectional STEM, EDS, and EELS map of a Cu on $MoS_2$ (a)-(c) and Cu on $SiO_2$ (d)-(f) device. The blue, red and pink line represents the signal of Cu, O and Si respectively in the cyan arrow direction. There is a C layer deposited on top of both devices during the sample preparation using focused ion beam (FIB) micromachining.

Before discussing the electrical performance, more details of the device structure should be analyzed. We carried out cross-sectional structure analysis and chemical mapping of a Cu on $MoS_2$ and a Cu on $SiO_2$ device by scanning transmission electron microscopy (STEM) in conjunction with energy dispersive X-ray spectroscopy (EDS) and electron energy loss spectroscopy (EELS). As it is shown in Figure 2(a)-(f), O signal was detected on the top surface of both devices because there was a $CuO_x$ layer formed due to Cu oxidation in air. Besides, there is another O signal on the $Cu/SiO_2$ interface as it is shown in Figure 2(e) and Figure 2(f). It is worth noting that in Figure 2(e), the second O signal, which is marked with green circle, appeared simultaneously with the Si signal, while the O and Si signals at the Cu upper surface marked in the black circle are different. Hence, the O signal belongs to $SiO_2$ rather than to $CuO_x$, so no $CuO_x$ is formed at $Cu/SiO_2$ or $Cu/MoS_2$ interfaces. The thickness of $CuO_x$ measured from the blue dashed lines in Figure 2(a)-(f) is around 2.5nm for both devices. We assume the $CuO_x$ thickness is the same for all the Cu films in both $Cu/SiO_2$ and $Cu/MoS_2$ structures. Besides, Atomic Force Microscope (AFM) was used to characterize the total thickness of $CuO_x/Cu$. After subtracting 2.5nm $CuO_x$ from the total thickness, we can identify the real thickness of Cu which is significantly important for the Cu electrical resistivity calculation from the measured resistance results in the electrical performance analysis.

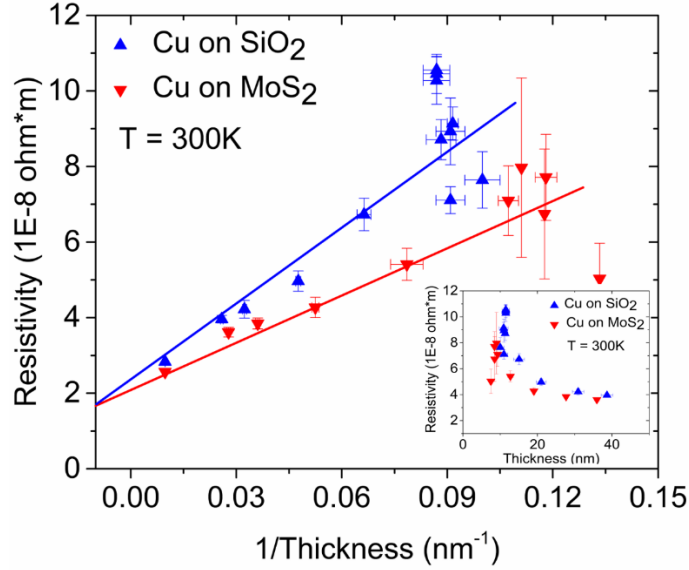

Figure 3. The resistivity of Cu on SiO$_2$ and Cu on MoS$_2$ as a function of the inverse of Cu film thickness at room temperature. The relationship between resistivity and Cu thickness is shown in the insert. Error bars capture the uncertainty in film thickness determination and resistivity calculation. The dots are experimental data and the solid lines are fitting results using FS analytical model.

Next, we will analyze the electrical performance of Cu thin films on different materials. Figure 3(a) shows the resistivity of Cu on SiO$_2$ and Cu on MoS$_2$ as a function of the inverse of Cu film thickness at room temperature. The corresponding relationship of resistivity and thickness is shown in the insert. The dots are experimental data, and the lines are theoretical fitting results with Fuchs–Sondheimer (FS) model. Each data point represents the averaged resistivity obtained from more than 10 individual devices with the same Cu thickness. Using four-probe method, we measured the resistance of Cu thin films in the geometry shown in Figure 1. During the measurement, we obtained similar resistance values with 400µA DC current and 10µA AC current, thus the Joule heating effect is negligible in our experiment. Comparing the resistivity of Cu on MoS$_2$ and SiO$_2$, we found: (1) the resistivity of Cu increases dramatically with the decrease of thickness regardless the underlying material. (2) when the thickness is larger than 100nm, the resistivity of Cu on MoS$_2$ and SiO$_2$ are similar. (3) for thinner Cu films, the resistivity on MoS$_2$ is smaller than that on SiO$_2$ and the thinner the Cu the larger the difference between these two cases. There are two reasons which may result in lower resistivity in Cu on MoS$_2$ case: (i) the semiconducting MoS$_2$ underneath Cu film works as a parallel electron transport channel which decreases the total resistivity of Cu; (ii) the electron scattering mechanism is different for Cu on different materials. To figure out which reason dominants, we measured gate dependence of the current in one Cu on MoS$_2$ device. It is well known that for a MoS$_2$ field effect transistor device, the on/off ratio should be 7-8 orders of magnitude [21-22]. If the MoS$_2$ plays an important role in the electron transportation, then the current through the device should be tuned dramatically by the gate voltage. However, the current only changes within 1% for -40V < V$_{gate}$ < 40V which means reason (i) should not be considered in our analysis. Thus, the resistivity difference should be related to different electron scattering in these two cases.

The contributions of surface scattering and grain boundary scattering to the total resistivity $\rho = \rho_S + \rho_G$ can be modeled by the Fuchs-Sondheimer (FS) equation (1) [23] and the Mayadas-Shatzkes (MS) equation (2) [24] respectively:

$$\rho_S = \rho_0[1 + \frac{3}{8}\frac{\Lambda_0}{T}(1-p)] \quad (1)$$

$$\rho_G = \rho_0[1 - \frac{3\alpha}{2} + 3\alpha^2 - 3\alpha^3 \ln(1 + \frac{1}{\alpha})]^{-1}, where\ \alpha = \frac{\lambda}{d_{grain}}(\frac{R}{1-R}) \quad (2)$$

Here, $\rho_0$ is the bulk resistivity of Cu, $\Lambda_0$ is the electron mean free path, T is the thickness of the film, p is the specularity parameter ranging from 0 (completely diffuse) to 1 (specular scattering), $d_{grain}$ is the average grain size and R is the grain-boundary reflection coefficient. The experimental resistivity shown in Figure 3 is linear with the inverse of Cu film thickness. Since the trap states at the Cu-oxide interface perturbs the smooth surface potential of Cu [14], we assumed the electron scattering at the Cu/CuO$_x$ and Cu/SiO$_2$ interfaces is completely diffusive (p = 0) and fitted the experimental results of Cu on SiO$_2$ with FS equation (Equation 1) to obtain the values of Cu bulk resistivity and electron mean free path. The fitting result is the blue solid line shown in Figure 3 corresponds to $\rho_0 = 1.69*10^{-8}$ Ωm and $\rho_0\Lambda_0 = 1.99*10^{-15}$ Ωm$^2$. Since the fitted bulk Cu resistivity is the same with the reported bulk value, together with the linear relationship between the resistivity and 1/Thickness, we regard the surface scattering as the dominant contribution to the electrical resistivity and the effects of the grain boundary scattering is negligible. However, the FS model has its intrinsic limitations because it is based on two approximations that are not justified for small thickness. It is assumed that the electronic structure is as in bulk and the surface is smooth, so the surface scattering occurs only at the surface. Thus, the FS model does not include the contribution from roughness scattering, it is not adequate to describe thin films with surface roughness. This causes the deviation of the fitted $\rho_0\Lambda_0$ from the acknowledged value $6.6 * 10^{-16}$ Ωm$^2$ [9, 25].

Later studies have proposed other models to include the contribution of surface roughness to the electrical resistivity [8, 13, 26-27]. Among them, the extended FS model [13] and the power law model developed by T. Zhou et al [26] claimed that the resistivity contribution from surface scattering is still proportional to 1/Thickness which coincides with our experimental results. They include a numerical factor $\alpha$ equal to or larger than 1 which is related to the roughness to the second term of Equation 1:

$$\rho = \rho_0[1 + \frac{3}{8}\frac{\alpha\Lambda_0}{T}(1-p)] \quad (3)$$

For Cu/SiO$_2$ devices, we used p = 0, $\rho_0\Lambda_0 = 6.6 * 10^{-16}$ Ωm$^2$ and fitted the experimental results with Equation 3. The fitting results are $\rho_0 = 1.69*10^{-8}$ and $\alpha = 3.02$. According to atomic force microscopy results, the surface roughness of Cu on MoS$_2$ is similar with Cu on SiO$_2$, thus $\alpha$ is the same for both cases. With all the parameters obtained from Cu on SiO$_2$ devices, we fitted the experimental results of Cu on MoS$_2$ with Equation 3 as the solid red line shown in Figure 3 and obtained p = 0.39 for the Cu/MoS$_2$ interface.

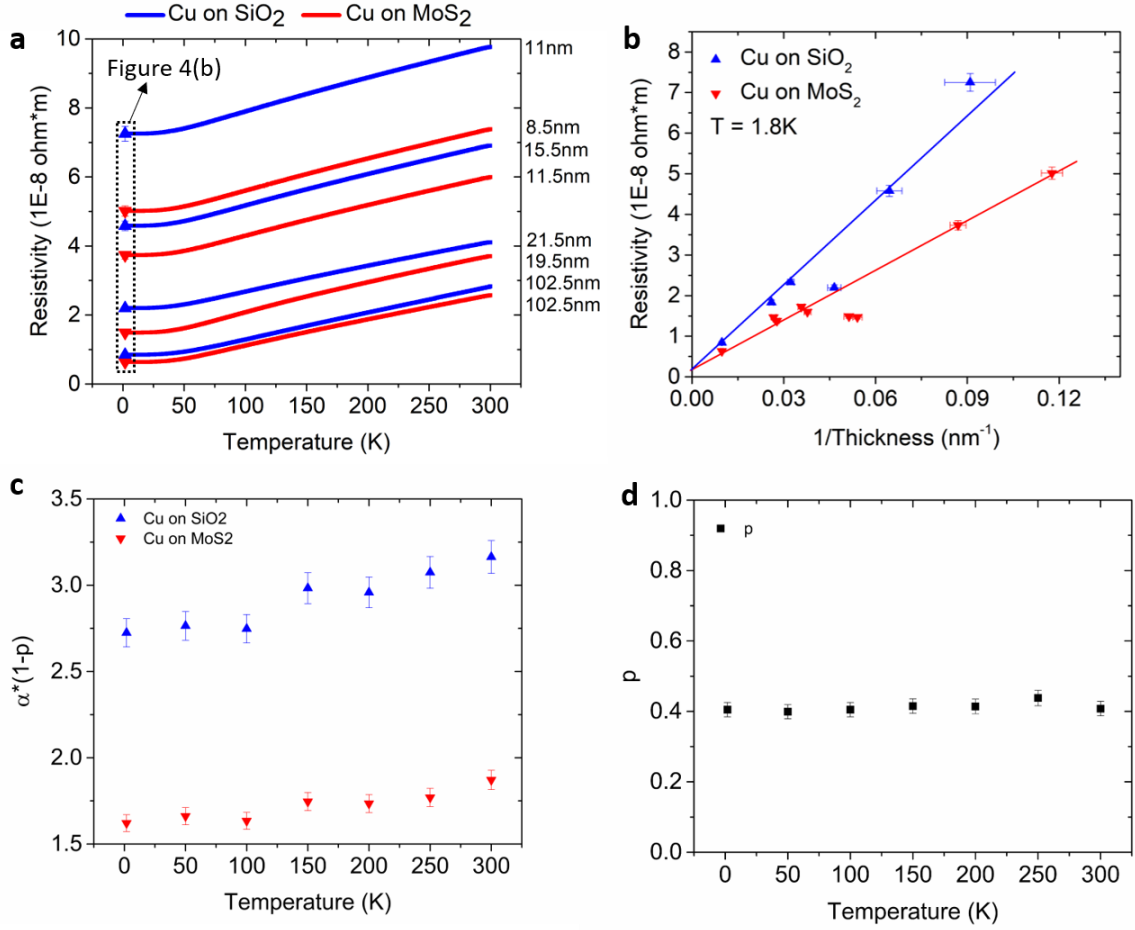

Figure 4. (a) The resistivity of Cu on $SiO_2$ and Cu on $MoS_2$ as a function of temperature for different Cu film thicknesses. (b) The resistivity of Cu on $SiO_2$ and Cu on $MoS_2$ as a function of the inverse of Cu film thickness at 1.8K shown in the dashed box in (a). The dots are experimental data and the solid lines are fitting results of FS analytical model. (c) The calculated $\alpha*(1-p)$ as a function of temperature of Cu on $SiO_2$ and Cu on $MoS_2$ extracted from (b) at different temperatures. (d) The calculated specularity parameter p as a function of temperature of Cu on $MoS_2$ from (c). Error bars capture the uncertainty in numerical calculations and film thickness determination.

In addition, we studied the temperature dependence of the electron scattering in Cu thin films. Figure 4(a) shows the experimental resistivity of Cu on $MoS_2$ (red curves) and Cu on $SiO_2$ (blue curves) as a function of temperature ranging from 300K to 1.8K for Cu films with different thickness. The measurement was carried out in a physical property measurement system (PPMS) with 10μA AC current. Each curve represents one set of experimental data obtained from one device. When 50K < T < 300K, the resistivity is linear with temperature and when the temperature is lower than 30K, the resistivity curves flatten out and reach constant residual resistivities [8, 28]. This is because at higher temperature phonon scattering dominants the total electoral resistivity. When the thermal energy becomes smaller than the phonon energy at low temperature, the phonon scattering is negligible and the contributions to the resistivity becomes temperature independent surface scattering, grain boundary scattering and impurity scattering. Moreover, within each set of the blue and red curves, thicker Cu films always show lower resistivity and for similar film thickness, the Cu resistivity on $MoS_2$ is lower than that on $SiO_2$ which agrees with the previous analysis.

Figure 4(b) is the residual resistivity of Cu in both cases as a function of 1/Thickness which is also labeled in the dashed box of Figure 4(a). The symbols are experimental data and the solid lines are eye guide linear relationship between Cu resistivity and 1/Thickness.

At low temperature, the FS model predicts $\rho \propto 1/[Tln(\Lambda_0/T)]$. However, this prediction cannot describe the experimental data correctly due to its intrinsic limitation: in the limit of high-purity films at low temperature, $\Lambda_0 \to \infty$, the FS model predicts a vanishing thin-film resistivity since surface scattering alone cannot relax carriers within the FS model [26]. Later studies reported by T. Zhou et al [26] have proposed another model which claims the resistivity contribution from surface scattering is temperature-independent and proportional to 1/Thickness to replace the FS model. Accordingly, we fitted the experimental results at different temperatures for Cu thin films on both $MoS_2$ and $SiO_2$ with Equation 3. Same as the analysis in Figure 3, we fitted the Cu on $SiO_2$ experimental data with p = 0 and $\rho_0\Lambda_0 = 6.6 * 10^{-16}$ $\Omega m^2$ which is temperature independent [29] and obtained the fitting values of α. Using the same α, we extracted p = 0.4 at the Cu/$MoS_2$ interface at 1.8K according to the slope difference of the blue and red solid lines. Same fitting was also done for experimental resistivity data at different temperatures. The fitted α*(1-p) for both Cu on $SiO_2$ and Cu on $MoS_2$ is shown in Figure 4(c). For Cu on $SiO_2$, the specularity is p=0 for all the temperatures because of the diffusive surface scattering. Thus, the blue points shown in Figure 4(c) represent α at different temperatures. The specularity of the Cu/$MoS_2$ interface extracted from the difference between the red and blue points is shown in Figure 4(d). Here, the fitting results show the specularity parameter at the Cu/$MoS_2$ interface is temperature independent with p ≈ 0.4 which means the Cu/$MoS_2$ has a temperature independent elastic surface scattering. Our experimental finding is consistent with the theoretical results reported by T. Zhou et. al [26]. The temperature dependent α*(1-p) in Figure 4(c) means the α is temperature dependent. Although α represent the surface roughness contribution to the total resistivity and should be independent of temperature, the slightly increase with temperature is understandable since as the increase of temperature, the trap charge density of states [30] at the Cu/$Cu_xO$, Cu/$SiO_2$ and Cu/$MoS_2$ interfaces might be increased slightly, and the surface scattering becomes more severe. As a result, the resistivity increases faster as a function of 1/Thickness at higher temperature. The fitting result in Ref [30] also show α varies with temperature, but we still need more studies to investigate.

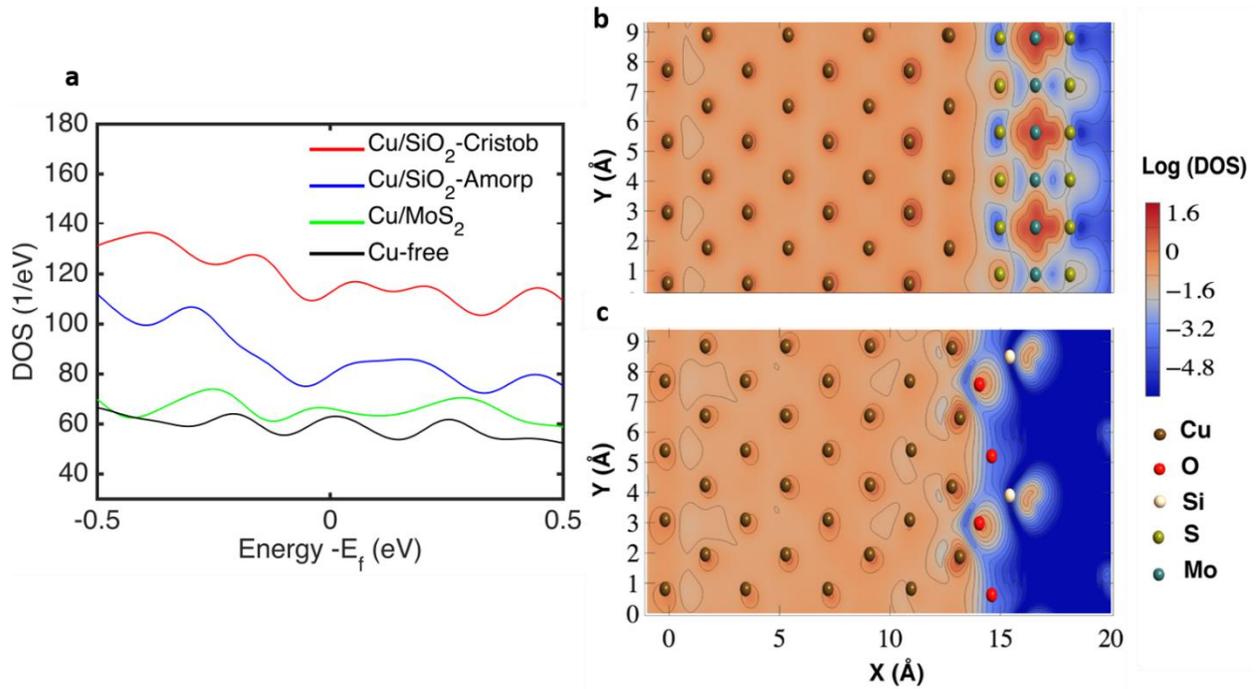

Figure 5. (a) Total DOS for Cu films with different interfaces. The red, blue, green and black corresponds to the simulated interface between Cu and cristobalite $SiO_2$, amorphous $SiO_2$, $MoS_2$ and a Cu interface with no passivated atoms respectively. Projected DOS at the interface of (b) Cu/$MoS_2$ and (c) Cu/amorphous $SiO_2$.

To explain the underlying mechanisms for the partial elastic surface scattering at the Cu/$MoS_2$ interface, we carried out first principle calculations based on the DFT to study the density of states at different Cu interfaces. The details of the computational simulation are discussed in the "Methods" section. As shown in Figure 5(a), the interface between Cu and $MoS_2$ has similar DOS with the free Cu surface, while the interfaces between Cu and cristobalite $SiO_2$ and amorphous $SiO_2$ have much higher DOS. Figure 5(b) and (c) are the projected DOS at the interface of Cu/$MoS_2$ and Cu/amorphous $SiO_2$ respectively. They show that the available states at these two interfaces are localized rather than continuous which means there are more trapping states at the Cu/amorphous $SiO_2$ interface than the Cu/$MoS_2$ interface. When electrons transport in thin Cu films, the probability to be trapped in the Cu/amorphous $SiO_2$ is much higher than that of the Cu/$MoS_2$ interface. Upon subsequent release, the trapped electrons have randomized momentum in the current flow direction. This explains why the Cu/$SiO_2$ heterostructures show higher resistivity. Our findings are in accordance with other works that demonstrate the interaction between $MoS_2$ and Cu is very weak [31] while the oxidation of the Cu surface or adsorption of foreign adatoms may cause perturbations to the Cu surface potential and effectively results in severe surface scattering [14, 32].

In summary, we studied the resistivity of thin Cu films on different materials and demonstrated two-dimensional $MoS_2$ can be used to enhance the electrical performance of Cu. With the scaling of the film thickness, the resistivity increases dramatically because of the diffusive surface scattering. However, by inserting $MoS_2$ under Cu, the resistivity can be decreased significantly due to the partial specular surface scattering at the Cu/$MoS_2$ interface. Our experimental results suggest a resistivity contribution from surface scattering on Cu surfaces is proportional to 1/Thickness at the temperature 1.8K-300K. From the analytical fitting results, we obtained a temperature independent specularity p ≈ 0.4 at Cu/$MoS_2$ interface. According to the DFT calculations, the higher resistivity in the Cu/SiO$_2$ heterostructure is caused by the higher density of localized states at the Cu/amorphous SiO$_2$ interface than the Cu/MoS$_2$ interface. Currently, only one surface of Cu thin film has been coated with MoS$_2$. If we could coat MoS$_2$ on all Cu surfaces, the resistivity of Cu can be reduced even further which is highly desirable for future generations of CMOS interconnects.

**Methods**

Few-layer MoS$_2$ flakes were exfoliated on Si/SiO$_2$ substrates followed by 200°C annealing for 5h in high vacuum. Subsequently, four probe test structures (4μm Length × 2μm Width) were fabricated on MoS$_2$ and SiO$_2$ surfaces with Cu thickness ranging from 8.5nm to 102.5nm using e-beam lithography, e-beam evaporation metal deposition and conventional lift-off. The thickness of Cu films was measured by an atomic force microscope (AFM) set-up.

**Computational Details**

To quantify the effect of the SiO$_2$ and MoS$_2$ interface over the copper atoms, first principles calculations were carried out by the density functional theory (DFT), using projector-argument waves (PAWs) as implemented in the VASP code [33]. In these calculations, the generalized-gradient approximation (GGA) with the Perdew-Burke-Ernzerhof (PBE) exchange-correlation functional were used [33]. In all the calculations, an energy cut-off of 500 eV with a convergence criterion of $10^{-8}$ eV and 0.1 eV Å$^{-1}$ for energies and forces, respectively, were used.

In this work, the Cu/SiO$_2$-Cristob configuration corresponds to the simulated interface between Cu and Cristobalite SiO$_2$. The configuration was constructed following the same process reported by T. Shan et al [34]. This interface was generated for copper oriented in the (001) direction which was matched to α-cristobalite (001). Based on the work in Ref 33, an oxygen terminated interface was chosen since this type of termination has the strongest adhesion energy between both materials. To reduce the strain effects at the edges of the interface, 8 atomic layers of each material were used in the x direction as show in Fig. 5 (b)-(c) and only four layers on each side were relaxed while the rest of the atoms were fixed during the ionic relaxation.

Making use of the structure previously described, Cu with amorphous SiO$_2$ was also studied. The amorphous silica used in the interface is prepared using the melt and quench method as suggested in Ref 35. This process was carried out with the ReaxFF potential modified for Cu/SiO$_2$ interface as reported in Ref 36 in the large-scale atomic/molecular massive parallel simulation (LAMMPS) [37]. During the molecular dynamics (MD) process, the copper atoms are fixed, and the atoms are melted from 300 K to 2000 K at a constant pressure for 200ps to ensure complete melting. Afterwards, the structure is quenched using a stepwise cooling scheme at a rate of 12.5mK/fs and the structure is equilibrated at 300 K for an additional 10ps and then relaxed in DFT making use of the same parameters used for Cu/SiO$_2$-Cristob configuration.

Finally, the Cu MoS$_2$ interface is obtained by straining the MoS$_2$ atoms to match the Cu interface and then the supercell is relaxed following the same process described for the SiO$_2$ cristobalite interface.

AUTHOR INFORMATION

**Corresponding Author**

*E-mail: shen168@purdue.edu

**Author contributions**

T.S. worked on the device fabrication, characterization and data analysis; D.V., K.W. and M.P. worked on the DTF calculations; Z.C., J.A. and G.K. analyzed the data and oversaw the planning and execution of the project; Q.W. and M.K. worked on the structure analysis; T.S. wrote the manuscript.

**Notes**

The authors declare no competing financial interest.

**ACKNOWLEDGMENTS**

The authors thank the staff at the Birck Nanotechnology center for their technical support. This work was in part supported by the STARnet center LEAST, a Semiconductor Research Corporation program sponsored by MARCO and DARPA. T. Shen is thankful for the help from Yuqi Zhu for discussions about data analysis.